\documentclass[preprint,authoryear,5p]{elsarticle-edit}

\usepackage[T1]{fontenc}
\usepackage{natbib} 
\usepackage{lineno} 
\usepackage{subfig}
\usepackage{makecell}
\usepackage{ulem}
\usepackage{adjustbox}
\usepackage{graphicx}
\usepackage{gensymb}

\title{Impacts and ejecta in natural granular material}

\author[a1]{Esteban Wright\corref{cor1}} 
\ead{ewrigh10@umd.edu}
\author[a2]{Emau Argueta}
\ead{esanch91@montgomerycolllege.edu}
\author[a1]{Wolfgang Losert}
\ead{wlosert@umd.edu}

\address[a1]{Institute for Physical Science and Technology, University of Maryland, College Park, MD 20742, USA}
\address[a2]{Department of Mathematics, Montgomery College,51 Mannakee St Rockville, MD 20850}

\cortext[cor1]{Corresponding author}

\begin{document}

\begin{abstract}
With laboratory experiments we investigate the ejecta of low-velocity ($\sim$m/s) impacts into multi-scale granular media and compare them against ejecta from impacts into mono-scale media.
Impacts are into a 50 cm diameter galvanized washtub filled with fine sand that has larger diameter gravel buried below the surface is filmed with two high-speed cameras.
The resulting ejecta curtain consists mainly of fine sand, and has a complex asymmetric structure that depends on the location and interaction of the ejecta with the larger gravel grains mixed into the sand.
To characterize the highly heterogeneous ejecta curtain we combine three analysis techniques:
Particle tracking measures the ejecta velocities and ejecta angles best in low density regions, while particle image velocimetry (PIV) elucidates average motion in dense regions, and histogram of oriented gradients (HOG) which  captures directions of motion against a patterned background.
We find significant asymmetries in the multi-scale ejecta's velocity distributions and ejection angles compared to the symmetry seen in the ejecta from impacts into mono-scale media.
Our experiments show that larger grains under the surface impede and direct ejecta along preferential paths during the impact process.
\end{abstract}


\maketitle


\section{Introduction}

\begin{figure}[h!]
    \centering
    \includegraphics[page=1, width=1.\columnwidth]{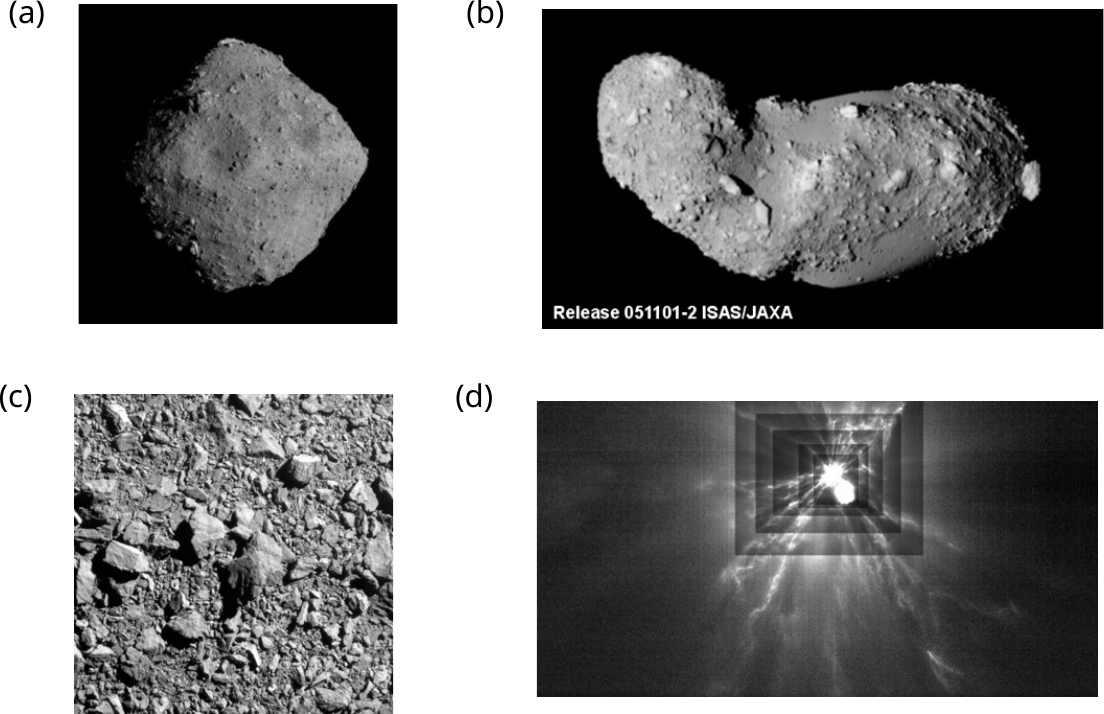}
    \caption{Images of asteroid surfaces showing material of several size-scales.
    The surface contains a broad size distribution of non-spherical granular material.
    Image of rubble-pile asteroids (a) 162173 Ryugu and 
    (b) 25143 Itokawa (JAXA, University of Tokyo \& collaborators). 
    (c) Surface of Dimorphos, target of the kinetic impact test mission DART, just before impact by the spacecraft (NASA/Johns Hopkins APL).
    (d) Image from LICIA cube post-impact by DART.
    The filamentary structure of the ejecta is clearly visible.
    Each square in the image has its contrast increased to highlight the complex structure distant from the binary system (ASI/NASA/APL).
    }
    \label{fig:figure_asteroids}
\end{figure}

Asteroid surfaces are polydisperse granular media and so are poorly approximated by experiments into a monodisperse medium. 
Missions to asteroids such as (25143) Itokawa, (101955) Bennu, and (162173) Ryugu have shown that rubble pile asteroids have diverse and heterogeneous surfaces \citep{veverka01, Lauretta_2019, DellaGiustina19, Michikami_2019, miyamoto07}.
Measuring the size distribution of boulders is a common way to characterize an asteroid's surface, given as a power-law in boulder size.
For example, the power-law slope for Bennu was measured to be -2.9 $\pm$ 0.3 \citep{DellaGiustina19}, and -2.65 $\pm$ 0.05 for Ryugu \citep{Michikami_2019}.
Figure \ref{fig:figure_asteroids} show surface images of the rubble-pile asteroids (a) Ryugu, (b) Itokawa, and (c) the asteroid-moon Dimorphos taken by JAXA's Hyabusa2 and NASA's DART spacecrafts, respectively.
For Ryugu's and Itokawa's surfaces, there is a wide range of material sizes with boulders measuring 10's of meters to fine grained ponds made of centimeter sized grains.
There is little reason to assume that the surface polydisperity should not continue below the surface, though the power law relation of the size distribution may change with depth \citep{Michikami_2019}.

The DART spacecraft impacted the asteroid-moon Dimorphos with a speed of about 6 km/s on September 22, 2022.
The impact produced a significant amount of ejected material from Dimorphos ($\sim 10^7$ kg) with a complicated filamentary structure as shown in Figure \ref{fig:figure_asteroids}(d) \citep{ferrari2025}.
This structure was the result of the polydisperity of material on the surface and sub-surface.

JAXA's Hyabusa2 spacecraft on its mission to study the rubble pile asteroid 162173 Ryugu carried on board the Small Carry-on Impactor (SCI) to impact the Ryugu's surface with an impact speed of $\sim$2 km/s.
The purpose of the impact experiment was to indirectly study the mechanical properties of the surface and test the dynamics of an in-situ impact.
Observations of pre- and post impact surface showed that boulders were excavated during the SCI impact on Ryugu and formed a 16 m diameter crater \citep{Arakawa_2020}. 
In proximity but outside the crater rim some boulders moved as a result of crater excavation. 
The ejecta curtain produced also had a complex structure with signs of obstruction of ejected material by buried boulders \citep{Honda_2021}.

The dynamics of impacts into granular media can be sensitive to particle size.
Lab studies of low velocity projectiles that ricochet find that deflection angle and velocity is sensitive to particle size \citep{Wright_2020b}.
With the exception of \citet{Ormo_2022},
few laboratory studies have focused on  
impacts into polydisperse granular media.
\citet{Ormo_2022} found that boulders were ejected at shallower impact angles compared to smaller sized sand, allowing them to pass through the sand ejecta curtain and land further from the impact site. 
\citet{Ormo_2022} did not track particle trajectories. 
When a polydisperse material is lofted by a subsurface pulse, \citep{wright19} found ejecta velocity and trajectories are independent of particle size.
However, boulders wind up on the surface due to ballistic sorting upon landing.

We go beyond these recent studies by studying the ejecta trajectories of normal impacts into polydisperse granular media.
We also characterize the velocity distribution and shape of the produced ejecta curtains. 

\section{Experimental Setup}\

\begin{figure}[]
    \centering
    {
    \includegraphics[page=1, width=.9\columnwidth]{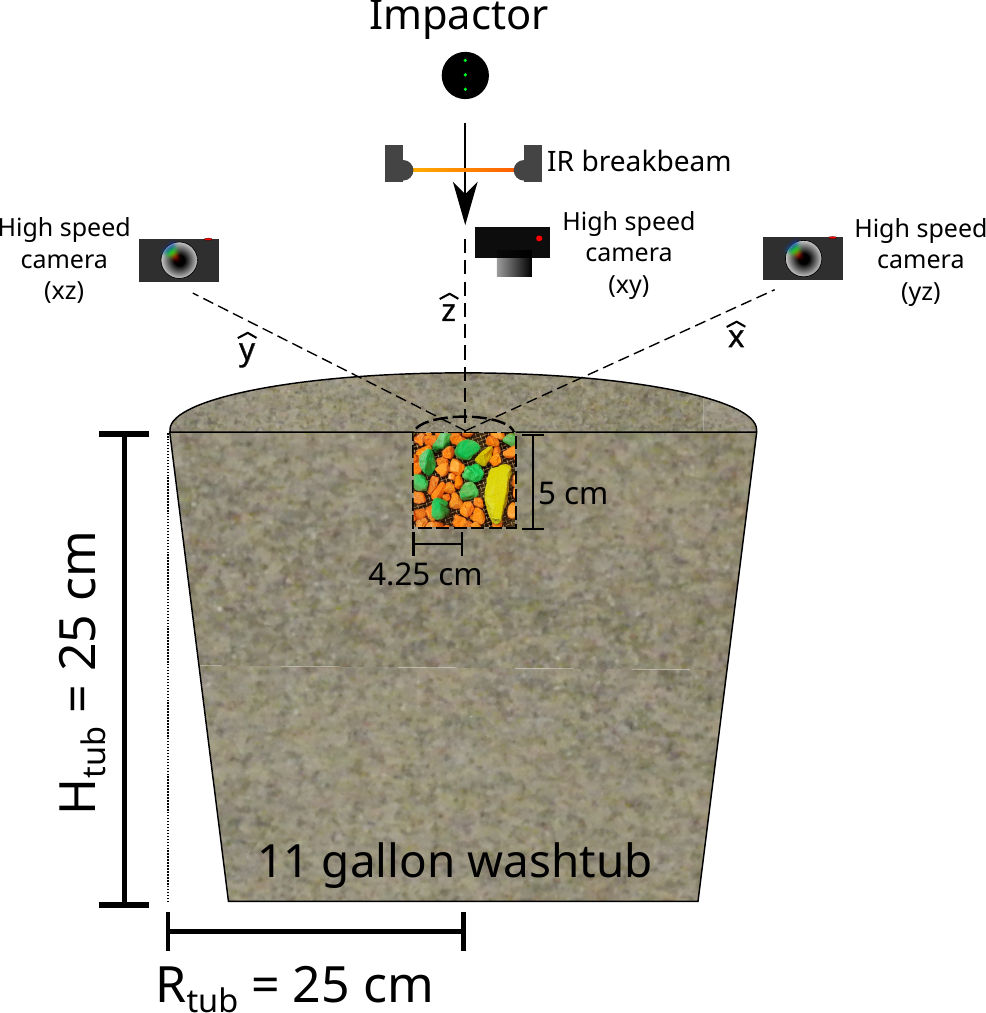}
    }
    \caption{Cartoon of the cross section of the washtub with a polydisperse mixture of granular material embedded in monodisperse sand. 
    The polydisperse mixture occupies a cylindrical region that extends 5 cm below the surface and is 9 cm in diameter. 
    A glass marble impactor is allowed to fall from a height of $z_{\rm drop} \sim 198$ cm to impact the granular substrate at normal incidence with an impact velocity of $v_{\rm impact} \sim  6.24$ m/s. 
    The impact is filmed with three Krontech Chronos 2.1 high speed cameras: one from above the impact and another two from the side located orthogonal to each other. 
    The falling impactor passes through an IR break-beam sensor which triggers the cameras to begin recording.
    Each camera is aligned with one of the system's coordinate axes filming a different plane of the ejecta curtain.
    The composition of the polydiperse mixture is given in Table \ref{tab:physical_values}, and its size distribution is given in \ref{fig:poly_mix_sfd}.
    For comparison, impact experiments were also done in only sand with the polydisperse mixture removed.
    }
    \label{fig:exp_setup} 
\end{figure}

We perform low velocity normal impacts into a polydisperse mixture of granular material and into monodisperse sand as a comparison. 
The granular media is held in an 11-gallon (41.6 liter) galavanized washtub that is predominately filled with monodisperse playground sand that is surrounding a cylindrical region of a polydisperse mixture as shown in Figure \ref{fig:exp_setup}.
The radius of the washtub $R_{\rm tub}$ is 25.1 cm and has a depth $H_{\rm tub}$ of 25.0 cm.
The tub is sufficiently large that we can ignore any effects of the container boundary on our experiments.

Our impactor is a spherical glass marble with radius $R_{\rm imp}$ = 1.75 cm and density $\rho_{\rm imp}$ of 2.77 $\rm g/cm^3$ .
The projectile is released from a drop height $z_{\rm drop}$ of 198 cm.
When the projectile is released it is only under the influence of gravity and therefore has an impact velocity given by $v_{\rm impact} = \sqrt{2gz_{\rm drop}} \approx$ 6.2 m/s.

\begin{figure}[]
    \centering
    \includegraphics[page=1,width=1.\columnwidth]{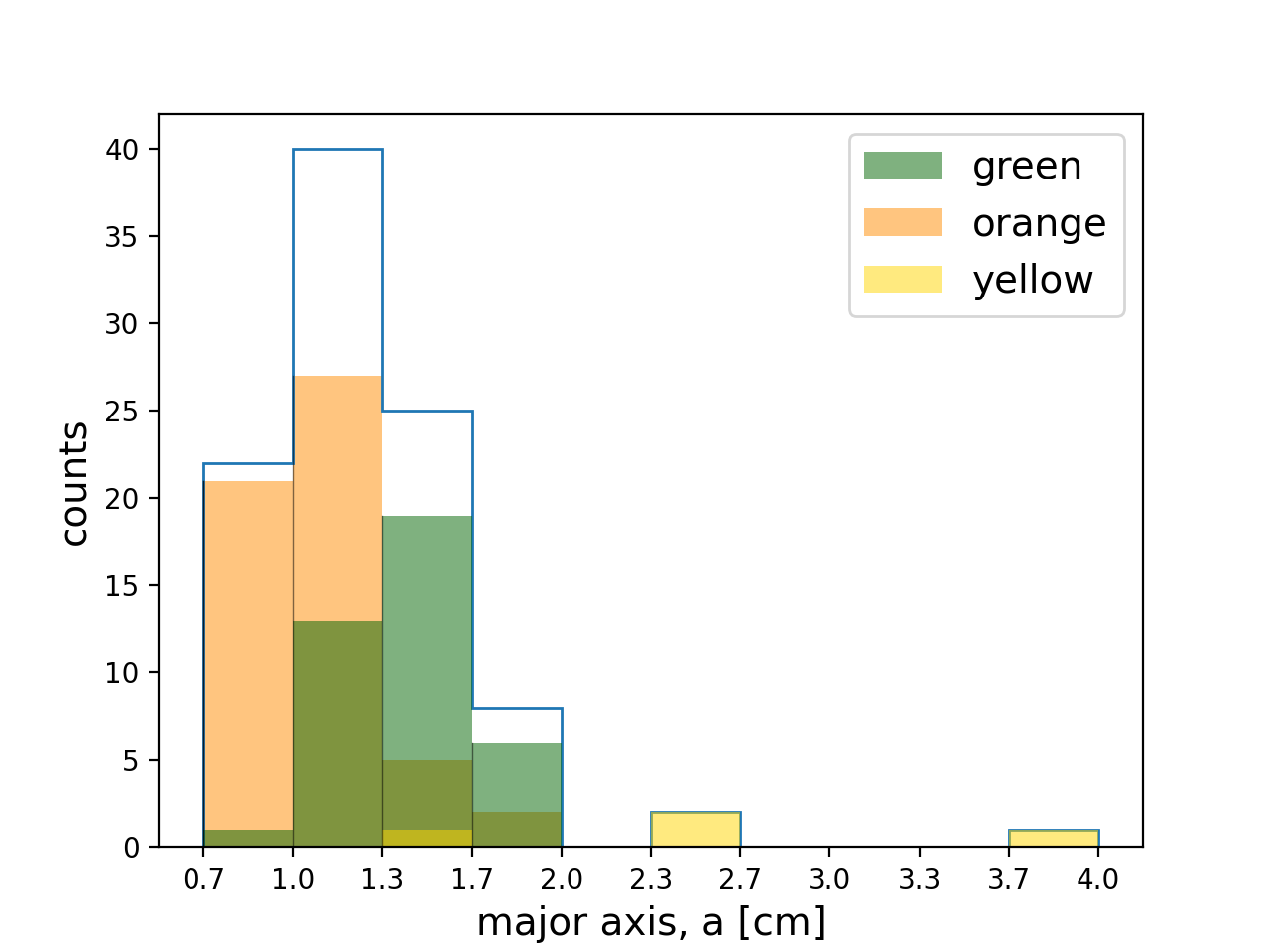}
    \caption{Size frequency distribution of the major axis of the polydisperse mixture embedded in the center of the experimental washtub.
    The color of the bars corresponds to the color of the grains. 
    The blue line is the total number of large grains in that given bin range.
    The composition of the polydisperse mixture is give in Table \ref{tab:physical_values}.
    }
    \label{fig:poly_mix_sfd}
\end{figure}

The polydisperse mixture was limited to a cylindrical region centered in the washtub with a radius $R_{\rm poly}$ of 4.25 cm and extended to a depth $h_{\rm poly}$ of 5.0 cm.
Our sand grains are nearly spherical with size $a \sim$ 0.5 mm and density $\rho_{\rm sand}$ = 1.5 $\rm g/cm^3$.
Larger grain sizes in the mixture ranged from $a \sim$ 1.0 -- 4.0 cm and painted with fluorescent paint.
Yellow material constituted the largest grain size with long axis lengths of $a \sim$ 2.5 -- 4 cm. 
Green rock had a size of $a \sim$ 1.0 -- 2.0 cm, and orange gravel with a size of $a \sim$ 0.5 -- 1.5 cm. 
The polydisperse mixture also included the playground sand which fill the voids between the large grains.
Alternatively, our impactor to grain size ratio spans $\sim$ 1 -- 80.
Figure \ref{fig:poly_mix_sfd} shows the size frequency distribution of each grain size in the polydisperse mixture.
The colors of the bars corresponds to the color of the large painted grains, and the blue line denotes the total number of grains of size $a$.
The percent composition, by weight, of the polydisperse mixture was 2\% yellow, 7\% green, 10\% orange, and 81\% sand. 
The bottom of Table \ref{tab:physical_values} has a list of experimental values and polydisperse mixture composition.

\begin{figure}[]
    \centering
    {
    \includegraphics[page=1, width=1.\columnwidth]{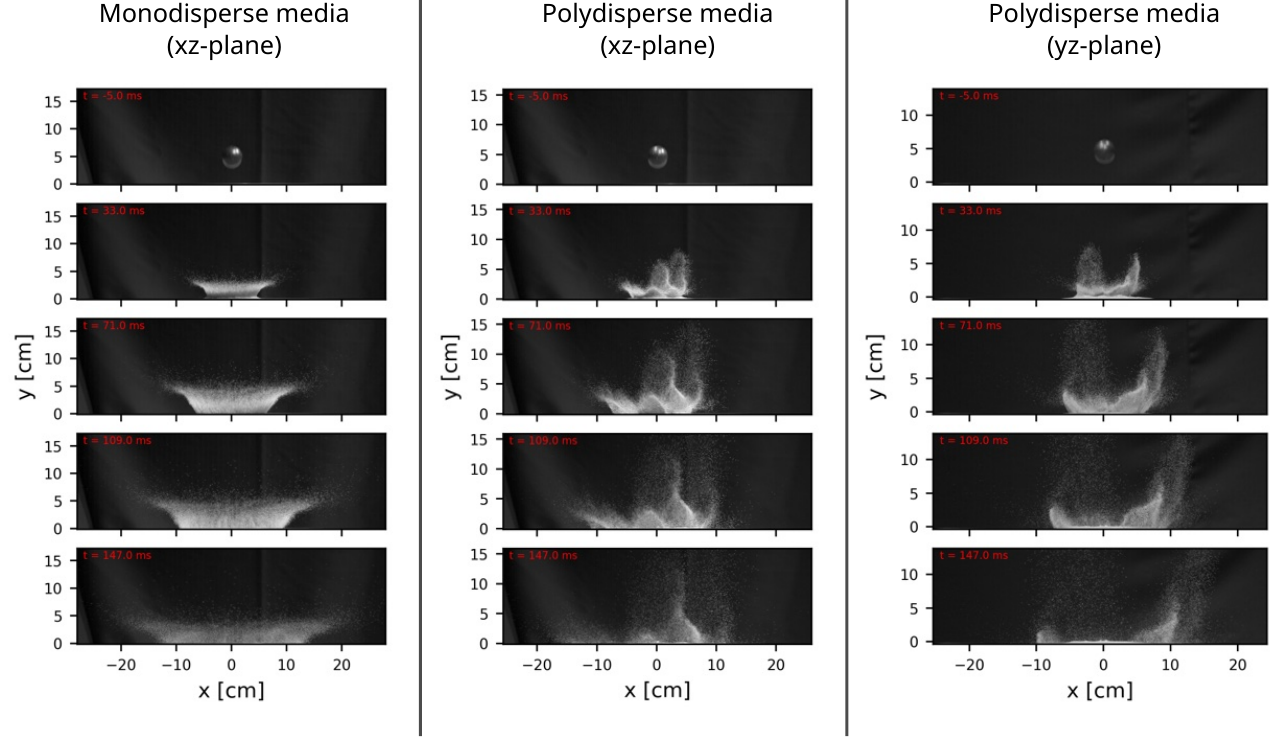}
    }
    \caption{High-speed video frames from an impact into the monodisperse granular substrate (left column), and an impact into the polydisperse granular mixture from the two side cameras in the middle and right columns.
    Videos were recorded at 1000 frames per second.
    Each frame is separated by 38 ms.
    The marble impactor has a diameter of 3.5 cm.
    }
    \label{fig:postcards} 
\end{figure}

We filmed the impacts using three Krontech Chronos 2.1 high-speed cameras.
One camera was positioned above the impact site looking down onto the surface of the granular material (i.e. the xy-coordinate plane).
The second and third high-speed cameras were positioned orthogonal to each other and to the sides of the washtub to film a side view of the impact (i.e. the yz- and xz-coordinate planes).
See Figure \ref{fig:exp_setup} for a schematic of the high-speed camera positions in relation to the washtub.
Our high speed videos were filmed at a framerate of 1000 frames per second (fps).
The high-speed cameras are connected to an infra-red break-beam sensor which triggers them to begin recording when the impactor passes through the beam.
Figure \ref{fig:postcards} shows frames from the high-speed videos located at the side of an impact into our granular media.
The left column is an impact into the monodisperse media, and the middle and right columns are of an impact into the polydisperse media from the two cameras on the side. 
Only one side camera of the impact into monodisperse media is shown since the ejecta curtain is the same from both viewpoints.
The high-speed videos are used to analyze the ejecta spatial and velocity distributions and the shape of the ejecta curtain.
A list of experimental high speed videos is given in Table \ref{tab:video_list} along with additional details about each video.

Between impact experiments the polydipserse mixture was mildly agitated to remove any packing that may have resulted from the impact.
The agitation means that the locations of individual large painted grains in the cylindrical region will be different between impacts.
This process of not removing the polydisperse mixture between impacts means that the produced ejecta curtain will be different for each experiment as the impact ejecta is affected by the subsurface material in the impact area.

\begin{table*}[]
    \centering
    \begin{tabular}{ccccc}
        Video & Viewpoint & Granular setup & Figure\\

    \hline
        Video 1 & above (xy) & monodisperse control & \ref{fig:piv_postcards}\\
        Video 2 & side (yz) & monodisperse control &  ---\\
        Video 3 & side (xz) & monodisperse control & \ref{fig:trackpy_results}, \ref{fig:HOG_postcards}\\
        
    \hline
        Video 4 & above (xy) & polydisperse mix & \ref{fig:piv_postcards}\\
        Video 5 & side (yz) & polydisperse mix & \ref{fig:trackpy_results}, \ref{fig:HOG_postcards}\\
        Video 6 & side (xz) & polydisperse mix & \ref{fig:trackpy_results}, \ref{fig:HOG_postcards}\\
    \hline

    \end{tabular}
    \caption{Experimental high speed video list.
    }
    \label{tab:video_list}
\end{table*}

\begin{table}[]
    \centering
    \begin{tabular}{lll}
    \hline
        Washtub radius & $R_{\rm tub}$ & 25.1 cm \\
        Washtub height & $H_{\rm tub}$ & 25.0 cm \\
        Polydisperse column radius & $R_{\rm poly}$ & 4.25 cm \\
        Polydisperse column depth & $h_{\rm poly}$ & 5.0 cm \\
        Substrate density & $\rho_{\rm sand}$ & 1.5 $\rm g/cm^{3}$ \\
        Grain size & $a$ & \\
        \hline
        Impactor radius & $R_{\rm imp}$ & 1.75 cm\\
        Impactor mass (sphere) & $M_{\rm imp}$ & 62.23 g \\
        Impactor density & $\rho_{\rm imp}$ & 2.77 $\rm g/cm^3$ \\
        Drop height & $z_{\rm drop}$ & 
        198 cm \\
        Impact velocity & $v_{\rm impact}$  & 6.24 m/s \\
         & $\approx \sqrt{2gz_{\rm drop}}$ & \\
    
    \hline
        Polydisperse column mixture & Yellow & 2\% \\
        (\% Weight)                 & Green  & 7\% \\
                                    & Orange & 10\% \\
                                    & Sand   & 81\% \\
    \hline
    \end{tabular}
    \caption{Experiment physical values and nomenclature. 
    }
    \label{tab:physical_values}
\end{table}

\section{Ejecta Asymmetry and trajectories}

To study the ejecta curtain 
we use several image analysis techniques on our high-speed videos to analyze the ejecta spatial and velocity distributions and the shape of the ejecta curtain.
The analyses are particle tracking with trackpy, particle image velocimetry (PIV), and the histogram of oriented gradients (HOG).

We track the sand grains in our high-speed videos using the python package trackpy \citep{trackpy}.
Trackpy uses the Crocker-Grier algorithm to identify and track blob-like features in a video.
We track the small grain ejecta from our two high-speed cameras positioned to the sides of the impact site, perpendicular to each other. 

\begin{figure*}[h!]
  \centering
  \includegraphics[page=1, width=2.\columnwidth]{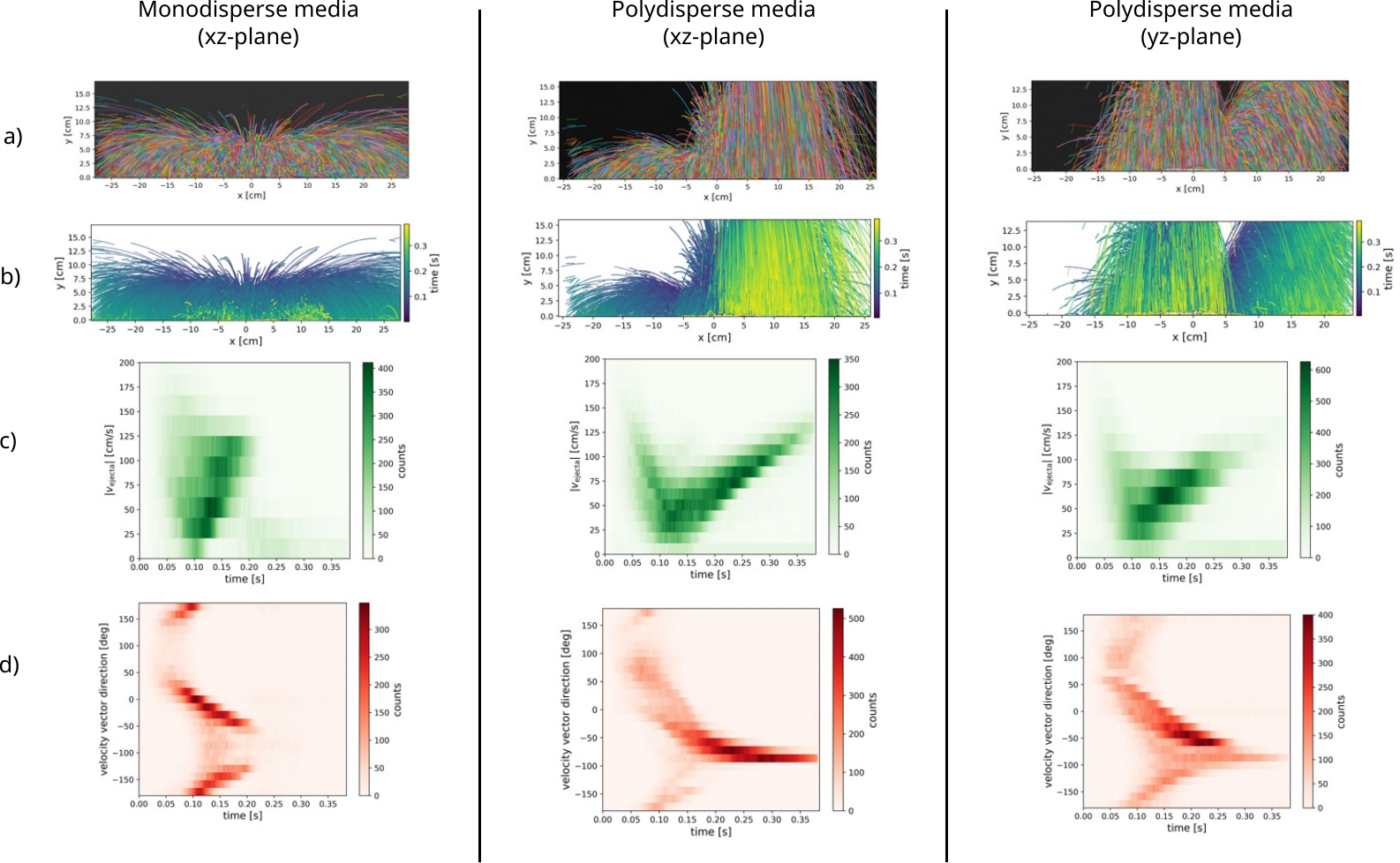}
    \caption{
    Trajectories of tracked sand grains in the ejecta curtain of a normal impact into a monodisperse (left column) and polydisperse mixture (middle and right columns).
    For the impact into polydisperse media tracking from both of the side high-speed cameras are shown (i.e. the xz- and yz-planes).
    The blue channel of our high speed videos were used to isolate the sand grains from the larger sized material.    
    (a) All ejecta trajectories found using the python package trackpy \citep{trackpy}. Each color corresponds to a unique particle trajectory. 
    (b) The same trajectories as shown in (a) but colored by time from impact ($t=0$).
    (c) Evolution of the ejecta velocity magnitude distributions over time.
    The horizontal axis is time in milliseconds and ejecta velocity on the vertical axis in cm/s.
    (d) Evolution of the ejecta velocity direction distributions over time.
    Ejecta velocity direction is measured relative to the horizontal, in degrees.
    Compared to the ejecta from an impact into the monodisperse media, ejecta from the polydisperse media is significantly more asymmetrical in velocity magnitudes and over time.
    }
    \label{fig:trackpy_results}
\end{figure*}

Figure \ref{fig:trackpy_results} shows the tracking results from both of the side cameras (xz- and yz-planes in Figure \ref{fig:exp_setup}).
The left column is the tracking results from an impact into the monodisperse sand, whereas the middle and right columns are the results from impacts into the polydisperse mixture from the xz- and yz-planes of the system, respectively.
Only one side view of the monodisperse impact is shown since the ejecta curtain structurally and dynamically is the same in both camera views, thus the result presented for the xz-plane is representative of the yz-plane tracking.

Figure \ref{fig:trackpy_results}(a) shows the ejecta trajectories with each unique particle trajectory uniquely colored.
Figure \ref{fig:trackpy_results}(b) is similar to (a) but the ejecta particle trajectories are colored by time from impact (t=0), in seconds.
Row (c) of \ref{fig:trackpy_results} show the evolution of the distributions of the ejecta velocity magnitudes $|v_{\rm ejecta}|$ measured from the tracked trajectories.
Similarly, \ref{fig:trackpy_results}(d) shows the distribution of the ejecta velocity direction with respect to the horizontal is shown in time.
Compared to the ejecta from an impact into the monodisperse media, ejecta from the polydisperse media is significantly more asymmetry in physical space 
Ejected particles from impacts into monodisperse media have more ballistic trajectories and have similar flight times.

The ejecta velocity magnitude distributions over time (Figure \ref{fig:trackpy_results}(c)) show  show key differences in the impact ejecta between the two granular media.
The velocity magnitude distributions of the monodisperse impact is clustered in time with the distributions keeping the same general shape and median value, though some broadening in the distribution occurs later in time.
Conversely, the velocity magnitude distributions of the polydisperse impacts are not symmetric, displaying a clear change in the median velocity with increasing time.
The ejecta velocities reach their lowest values ($\sim$ 50 cm/s) about 0.12 seconds after impact.
The late temporal asymmetry in the velocity magnitude distributions can be understood by seeing that sand on the right side of the ejecta curtain in the xz-camera plane (middle column) is falling downward about 0.2 s later than material on the left side.
From the yz-camera plane, two distinct lobes are seen on either side of the impact point with similar time of flights for the sand in the ejecta curtain.

The ejecta velocity direction distributions in time are given in Figure \ref{fig:trackpy_results}(d).
Impacts into the monodisperse media have a very symmetric distribution in velocity direction.
There are two main regions corresponding to tracks on the left and right of the impact site.
The distributions in the polydisperse media in both camera views have weak bimodal distributions early in time but a strong signal late in time emerges from ejecta with long time-of-flights falling downward (--90\textdegree) back onto the surface.

Particle image velocimetry is a technique used to measure the flow field of a fluid or fluid-like system.
We use OpenPIV \citep{openpiv}, an open source python package for performing PIV analysis on our ejecta curtains using the high-speed videos filmed from above the impact site.
The advantage of using PIV over particle tracking for this viewpoint is the ejecta curtain does not stand out against the light colored background surface of the granular bed like in the side view videos (with a dark background).
PIV provides a measurement of the ejecta velocity field distributed azimuthally around the impact site in time. 
PIV is able to nicely capture the flow of the ejecta around the largest grains near the impact site which impede the travel of the smallest sand grains, shielding the ejecta 
resulting in azimuthal asymmetry in the ejecta curtain. 

\begin{figure*}[p!]
  \centering
  \includegraphics[page=1, width=1.5\columnwidth]{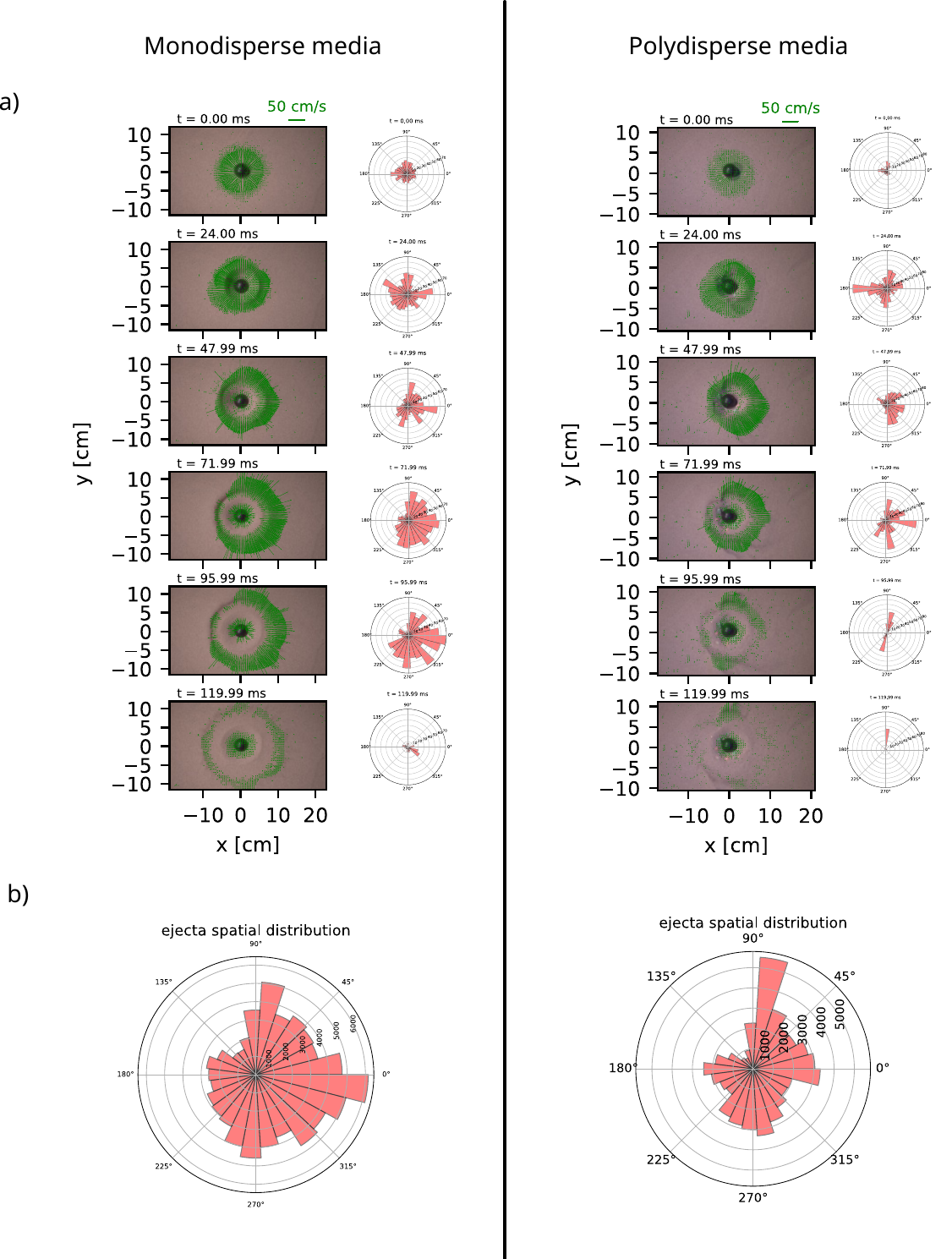}
\caption{
PIV results of an impact into a monodispaerse (left) and a buried polydisperse (right) granular system.
(a) The measured velocity field found using PIV.
Green arrows are velocity vectors to illustrate the flow filed of the ejecta curtain.
Each image is a frame from the high-speed video separated by 24 ms, with t=0 corresponding to the time of impact.
To the right of each frame is the azimuthal angular distribution of the velocity vectors about the impact site at each time step.
(b) The azimuthal angular distribution of the velocity vectors for the whole duration of the video.
Compared to the monodisperse media, the ejecta velocity field of the polydisperse media has clear azimuthal asymmetry around the impact site due to the interference of the ejecta from large grains buried below the surface.
}
\label{fig:piv_postcards}
\end{figure*}

Figure \ref{fig:piv_postcards} illustrates the results of our PIV analysis and the resulting flow field of the ejecta curtain.
The left and right columns in the figure are of the above camera view for an impact into the monodisperse and polydisperse granular media, respectively.
Figure \ref{fig:piv_postcards}(a) show sequential snapshots in time with each still frame separated in time by 24 ms.
The measured velocity vectors (green arrows) associated with the flow field of the ejecta curtain are overlayed on the associated frame. 
To the right of each still frame is the azimuthal distribution of velocity vectors about the impact site at that moment in time. 
Below the postcards, Figure \ref{fig:piv_postcards}(b) is the instantaneous velocity vector direction distribution for the entire impact process.
The monodisperse impact has a more uniform distribution compared to the distribution from the polydisperse impact.
The monodisperse distribution is shifted toward the 0\textdegree (right side) due to the ejecta curtain's shadow on the left.

Due to large sized grains buried below the impact site the ejecta curtain has a complicated shape.
The curtain has filamentary structure, varying density, asymmetrical velocities, and different launch angles.
To characterize this structure we perform a Histogram of Oriented Gradients (HOG) analysis on the video frames \citep{hog2005}. 
HOG works by calculating the intensity gradient and it's orientation for each pixel in an image, bins the pixels in a local region, and computes a weighted gradient vector for the binned region.
This process allows for characterizing the shape of the ejecta curtain.

\begin{figure*}[h!]
  \centering
  \includegraphics[page=1, width=2.\columnwidth]{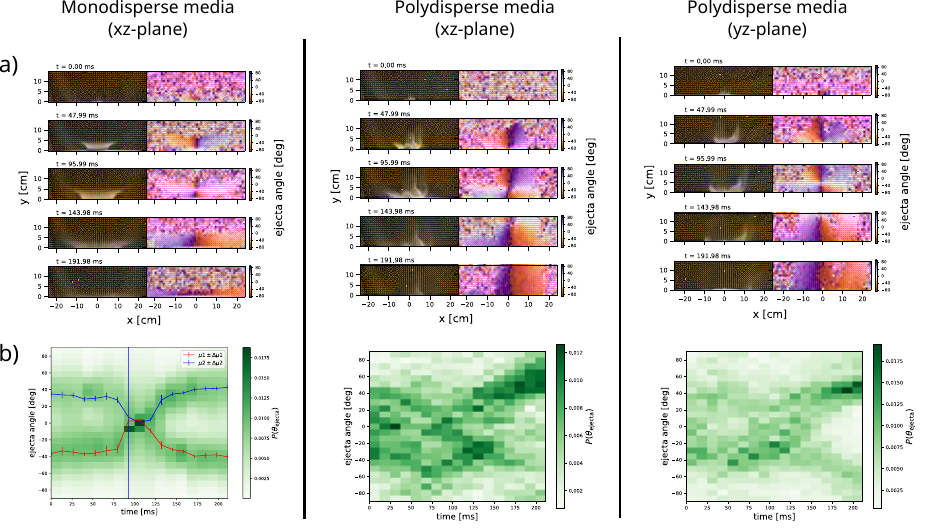}
\caption{
Histogram of Orientated gradients (HOG) of impact experiments into monodisperse (left) and polydisperse granular media (middle and right).
Two orthogonal camera planes are shown for the polydisperse impact.
a) Snapshots of our high-speed videos with each row being separated by a constant interval.
Times are marked in the upper left corner of each snapshot.
The left side of each column has the orientation of the gradient vectors in yellow overlayed on the corresponding video frame.
The right side of each column shows the angle of the gradient vector between $\pm$ 90\textdegree. 
Purple is for positive angles and orange for negative angles.
Angles are measured with respect to the horizontal axis.
b) Probability of ejecta angle $P(\theta_{\rm ejecta})$ in time.
For the monodisperse media, red and blue lines are plotted to show the mean ejecta angles $\mu$ of a bimodal gaussian model fitted to the distributions of the ejecta angles. 
The vertical lines are errors in $\mu$.
Two prominent peaks are clearly present early and later in time, with a convergence around 100 ms when the ejecta curtain reaches max height and begins to turn over and fall.
However, for the polydisperse media there is no underlying distribution to the ejecta angles, but regions of relative high-probability exist and change over time.
}
\label{fig:HOG_postcards}
\end{figure*}

Figure \ref{fig:HOG_postcards}(a) shows the result of the HOG analysis on the high-speed videos: the monodisperse xz-plane in the left column, and the polydisperse xz- and yz-planes in the middle and right columns, respectively.
Each row is a different video frame separated by 48 ms.
The left column in each panel is of the original video frame (in grayscale) overlayed with the resulting gradient orientations, in orange.
The right column of each panel is a diverging color map of the orientation angles with purple coloring being positive and orange being negative angles, ranging between $\pm$ 90\textdegree.
Regions of the curtain on either side of the impact site show a non-uniform spatial distribution in the gradient for the polydisperse impacts.
No such asymmetries are present in the ejecta curtain for impacts into the monodisperse media.

We plot the probability distributions of the ejecta gradient angles $P(\theta_{\rm ejecta})$ to show the evolution of the ejecta angles with time in Figure \ref{fig:HOG_postcards}(b).
For the monodisperse impact, the ejecta angle distributions had a bi-modal shape that evolved to a uni-modal shape centered at 0\textdegree around 100 ms from impact.
At this time the ejecta reaches maximum height and turns over, just before falling back to the surface.
As the ejecta is falling the bi-modal shape to the distribution returns as either side of the ejecta curtain regains the symmetry in angle.
We fit a bi-modal gaussian function to these distributions to measure a mean angle of each mode, $\mu_1$ and $\mu_2$, plotted as red and blue lines, respectively.
The associated errors are plotted as small vertical bars on the red and blue lines.

The polydisperse impacts did not have such a structured distribution in time.
Instead, they show distinct regions where certain angles more prevalent than others.
However, these regions shift as time progresses.
Since there was no underlying distribution to the data we did not attempt to fit a function to the data.

\section{Discussion}
We performed impact experiments of a glass sphere into monodisperse sand and a polydisperse granular medium.
Our experiments into the buried polydisperse mixture showed significant asymmetries in the ejecta curtain's spatial and velocity distributions, ejecta angle, and ejecta curtain shape.
We used three image analysis techniques, particle tracking, particle image velocimetry (PIV), and histograms of oriented gradients (HOG), on high-speed video recordings of the impacts.
We have outlined how large buried grains strongly affect the shape and timing of the ejecta curtain leading to dynamically distinct regions of the curtain.
Buried boulders can shield and/or slow the sand grains in the curtain and are not required to be on the surface before or during impact.

Impacts into the monodisperse media exhibit more spatial and ejecta velocity symmetry compared to an impact into the polydisperse media.
This symmetry lends itself well to measuring dynamical effects.
The polydisperse media, however, has much more variability in the dynamics of the resulting ejecta curtain due to the mixture being disturbed both by the impact itself, and by being agitated between experiments to break up any compaction that may have occurred during the impact.
This means that for each experiment the positions of the large grains change resulting in a unique ejecta curtain.
Future studies are needed to investigate broader statistical metrics and trends that may be independent of the underlying positions of the grains.

Post-impact, we find that boulders near the surface tend to be excavated as the sand is launched from the impact site.
We did not observe large grains being ejected with the sand in the ejecta curtain.
But experiments with higher energy impacts can eject large grains (see \cite{Ormo_2022} and \cite{ormo2023}).
In future experiments with higher energy impacts we can measure the velocity distribution of large grains to investigate if the dynamics differ compared to the small grains. 

The transient phase of the crater's formation is associated with material within the crater sliding down the crater walls to settle at a stable slope.
During this phase larger grains that were previously buried are excavated but are reburied from slumping material.
Other experiments resulted in larger grains being left uncovered. 
In most impacts, there are large grains that did not physically move, only smaller grained material on or near the boulders were displaced. 
This is interesting phenomena in of itself and the surface size frequency distribution (SFD) of the large grains can be measured as the number of impacts on the same system increases.
This could give insight into how the SFD evolves with repeated impacts.

Overall, our results indicate that impact experiments can reveal the size of buried boulders if impactors smaller than those boulders are chosen. Conceptually this is akin to the breakdown of treating the regolith as a fluid when it becomes comparable in size to the probe size. Thus experiments with arrays of impactors of a range of sizes may reveal the size of buried regolith from the onset of heterogeneity in ejecta directions, speed, and ejecta deposit shape. In turn, impacts with arrays of similar size impactors will reveal the heterogeneity in the underlying regolith. Thus rather than choosing impactors that enable a continuum mechanics analysis of their effect, we propose that smaller impactors, moving at slower speed may also reveal powerful insights into subsurface features.

\vskip 1 truein 
\section*{Acknowledgements} 
This material is based upon work supported by the National Science Foundation's MPS-Ascend Postdoctoral Research Fellowship under Award No. 2213195.

\vskip 0.3 truein

\bibliographystyle{elsarticle-harv}
\bibliography{refs}


\end{document}